# Quantum Chemistry on the time axis: electron correlations and rearrangements on femtosecond and attosecond scales

## Cleanthes A. Nicolaides

*Theoretical and Physical Chemistry Institute, National Hellenic Research Foundation, 48 Vasileos Constantinou Avenue, Athens 11635, Greece*     <u>caan@eie.gr</u>

Recent developments toward the production and laboratory use of pulses of high intensity, and/or of very high frequency, and/or of ultrashort duration, make possible experiments which can produce *time-resolved* data on ultrafast transformations involving 'motions' of electrons.

The formulation, quantitative understanding and prediction of related new phenomena entail the possibility of computing and applying solutions of the *many-electron time-dependent Schrödinger equation*, for arbitrary electronic structures, including the dominant effects of Rydberg series, of multiply excited states and of the multi-channel continuous spectrum. To this purpose, we have proposed and applied to many prototypical cases the *state-specific expansion approach* (SSEA). (Mercouris, Komninos and Nicolaides, Adv. Quantum Chem. **60**, 333 (2010)).

The paper explains briefly the SSEA, and outlines four of its applications to recently formulated problems concerning time-resolved electronic processes, where electron correlations are crucial. These are, 1) The time resolution of the decay of polyelectronic unstable states, 2) The excitation and decay of strongly correlating doubly excited states and atto-time-resolution of their geometries, 3) The time-resolved process of formation of the interference profiles of resonance states during the femtosecond photoionization of Helium and Aluminum, and 4) The relative time delay in the (2s,2p) photoionization of Neon by an attosecond pulse.

## 1. Prologue

The title of the plenary session of the 55[th] Sanibel symposium, February 15-20, 2015, where this paper was presented, was '*Attosecond spectroscopy and its demands on electronic structure and dynamics*'.

As regards 'attosecond spectroscopy', the field is at an early, but rapidly growing stage, especially with respect to its potential for scientific and technological applications. It belongs to a contemporary frontier of experimental science which is capable of producing *time-resolved* information about dynamics of ultrafast processes in atoms, molecules and condensed matter. In the framework of appropriate experimental techniques, the use of attosecond pulses, (1 a.u. = 24.2 as, 1 as = $10^{-18}$ sec), has created the possibility that, under certain conditions of excitation, the effects of the induced 'motion' and rearrangement of atomic and molecular electrons can be time-resolved on ultrashort scales. It is important to note that the frequencies of the attosecond pulses are in the extreme UV and beyond, and so their interaction with an atom inevitably renders the continuous spectrum physically relevant.

As regards the words 'electronic structure' and 'dynamics', and corresponding concepts, these are at the core of Quantum Chemistry (QC) and its dominant sector dealing with the computation of properties including the effects of electron correlation, with nonrelativistic or relativistic Hamiltonians. However, the fact is that the bulk of this activity in QC has traditionally focused on many-electron systems in terms of formalisms and computational methods that produce results on the *energy axis*. For example, either in the development of methods or in the plethora of applications, the overwhelming majority of publications in this area over many decades, report results and analysis concerning total energies and other properties of ground or of low-lying discrete stationary states (often as a function of geometry), radiative transition rates, etc. For small systems, such data are sometimes utilized in studies of reactions. Traditionally, the bulk of this activity has been focusing on the ground state, with the goal of being able to treat larger and larger systems. However, it should be added that there is also the rich domain of highly excited states and of the role of the continuous spectrum along the 'dimension' of increasing excitation, where things become more complex. For example, this is the case with resonance states, (also called autoionizing or autodetaching or Auger states, etc), in many-electron atoms and molecules, since their reliable treatment requires the combination of special many-electron methods with formalisms that account for the peculiarities of the continuous spectrum [1].

On the other hand, it is clear that the experimental capacity to *time-resolve* electronic processes in atoms and molecules (or condensed matter) within scales of tens or of hundreds of attoseconds or of a few femtoseconds, offers new challenges which fall in the realm of QC. Now, it is necessary to formulate and interpret problems of QC in terms of reliable solutions of the fundamental *many-electron time-dependent Schrödinger equation*, (METDSE), so as to be able to resolve and interpret the system's dynamics on the *time axis*. Evidently, if the theoretical formulation for the solution of a problem is correct, and if the discretization of time in the computation is on scales of attoseconds and femtoseconds, the scientific information which is obtained via the solution of the METDSE is equivalent to that obtained experimentally.

For most current and future problems involving the interaction of atomic (molecular) states with strong and/or ultrashort light pulses, it is necessary for the solution of the METDSE to be determined nonperturbatively. In the context of attosecond spectroscopy, the schemes of interest to *time-resolved many-electron physics* (TRMEP) involve time-dependent excitation, evolution and decay along electronic states, without of course excluding the possibility of studying effects of dynamics of nuclear motion as well. The principal requirement is for theory, (formalism, calculation and analysis), to be able to deal with arbitrary electronic structures of neutrals or of ions, and to account explicitly for the multichannel continuous spectrum, including all types of resonance (unstable) states that may be relevant to the phenomenon under study.

It is elements of this type of QC that the present paper addresses, by outlining the key features of the *state-specific expansion approach* (SSEA) [1, 2] to the solution of the METDSE, while referring to prototypical applications that demonstrated the possibility of calculating time-resolved information on electron rearrangements and correlations. Three examples are discussed briefly:

The first has to do with the theory and the quantitative description of the interplay between, on the one hand electronic structures and strong electron correlations, and, on the other hand, time-resolved excitation and/or decay of unstable many-electron states undergoing ultrafast electronic rearrangements caused by the system's Hamiltonian [3-6].

Publications [5,6] were the first to describe theoretical-computational work linking possible effects of time-resolved electron correlation and rearrangement in real systems to future experiments using attosecond pulses which had been generated in the laboratory for the first time in 2001. Specifically, they brought up the possibility of resolving on the time axis many-electron effects during excitation, state 'beating', electron rearrangement and autoionization involving states where electron correlation is strong. The proof of principle application dealt with the time-resolved dynamics of $^1P^o$ doubly excited states (DES) of Helium excited by femtosecond pulses from the $1s^2$ or the $1s2s\ ^1S$ states.

The second example has to do with the experimental-theoretical results which were published in 2010, on the TRMEP in photoionization by an attosecond pulse [7]. In that work, the theoretical treatment, which included the solution of the conceptual problem in terms of the energy derivative of phases of complex expansion coefficients (matrix elements) of scattering wavefunctions, was based on SSEA calculations, and showed reliably, in qualitative agreement with experiment, that the electron wavepackets representing the ejected electrons from the 2*s* and 2*p* subshells of Neon emerge with a relative *time delay*, of the order of a few attoseconds.

The third application has to do with the first ab initio demonstration of the time-resolved formation of electronic resonance states, as interference between bound and scattering components is taking place, with application to prototypical cases of DES of Helium and of inner-hole states of Aluminum [8,9]. In these cases, the process of formation of the resonance is completed in the range of femtoseconds. However, using an analytic formula valid for a weak attosecond excitation pulse [8], it can easily be predicted that for autoionization widths of isolated resonance states larger than about 2-2.5 eV, this process may be completed within hundreds of attoseconds.

## 2. The SSEA: Solution of problems of time-resolved many-electron physics (TRMEP)

### *2a. Introduction*

The problems that are generated by the interaction of atomic (molecular) states with strong and/or ultrashort light pulses and linked to TRMEP, have as their fundamental requirement the ab initio solution, $\Psi(t)$, of the appropriate for each case METDSE and the subsequent projection of $\Psi(t)$ on physically relevant stationary states.

By *many-electron* I mean atomic and molecular states other than the extensively treated singlet ground states of Helium and of Hydrogen molecule, for which the electronic structures are simple, and where direct one-electron photo-excitation, photo-ionization, (without excitation-ionization of the second electron), involves only one channel with a one-electron core. Perhaps needless to stress that, in almost all aspects of atomic and molecular time-independent or time-dependent physics, these two-electron states, (and certainly the one-electron H or $H_2^+$), do not exhibit the richness of possibilities and of complexities that characterizes the spectra and properties of polyelectronic systems, even though some interesting problems, often of just phenomenological nature, can be constructed and solved to a very good approximation via their investigation.

Furthermore, by *many-electron* it is also meant that the electronic structure in zero order need not be a single determinant. Even if near-degeneracy effects are absent, the construction of symmetry-adapted wavefunctions often requires open (sub)shell configurations to be written as linear superpositions of Slater determinants. For example, in atoms, such superpositions may require hundreds of determinants for just one symmetry-adapted configuration with open *d* and *f* subshells. Obviously, the situation becomes much more complex and demanding when the single configuration approximation breaks down in zero order, as it happens when heavy configurational mixing is present. In the framework of the '*state-specific theory*' [10, 11], the proposal and practice has been to search for a well-optimized zero-order wavefunction of one or of more configurations, where the orbitals are optimized for the state of interest. This is especially relevant to excited states having lower states of the same symmetry, including the large variety of unstable electronic states in the continuum ('shape', multiply excited, and inner-hole resonances) [11]. As we stated in a book reviewing developments in the Quantum Chemistry of excited states during the 1970s, '*The theory which we present in this lecture is <u>state - specific</u>. This means that the zeroth order function is radially optimized for each state and for each geometry separately. This allows consistent and systematic treatment of electron correlation in excited states. (In this way, electron correlation also becomes state - specific)*' (page 122 of [12]).

### *2b. The non-perturbative solution of the METDSE*

A popular technique which characterizes certain approaches to the solution of the METDSE, (thus far, only for a few special systems), is to propagate the solution on a grid of space-time points, ($x,t$), ($x$ represents collectively the coordinates chosen for each problem), using suitable formulations for the 'discretization' of the wavefunctions and of actual or the effective operators. When it comes to many-electron, closed-shell states, the zero order approximation in this context is the '*single active electron*' approximation [13]. Excepting one-and two-electron singlet states, going beyond the SAE approximation for arbitrary electronic structures interacting with strong and short electromagnetic pulses, the approach of using multidimensional space-time lattice integration schemes entails formal and computational difficulties that seem to be insurmountable.

On the other hand, in the case of one-and two-electron states, methods that employ ($x,t$) lattices and pursue the direct integration over the electron coordinates of the TDSE as a partial differential equation, normally implemented via techniques of expansion over a basis set, (e.g., bipolar spherical harmonics), have been developed and applied by various groups, albeit with formidable requirements of computer power. An indicative (incomplete) list of publications presenting formal and numerical results on two-electron problems is [14-18].

The nonperturbative methodology of the SSEA does not employ a ($x,t$) grid technique at any stage. It is based on an expansion over N-electron wavefunctions which are computed separately, following concepts and many-electron methods from the theory of electronic structures of ground and excited discrete states, of resonance states, and of purely scattering states [2,10,11]. Hence, it is applicable to both closed- and open-shell states and to corresponding time-dependent dynamics. Its salient features are described in the following paragraphs.

In the early 1990s, we concluded that, the fundamental quantum mechanical principle regarding the expansion for $\Psi(t)$ in terms of stationary states is the solid ground for developing a rigorous, general, practical and physically transparent approach to the ab initio non-perturbative solution of the METDSE for a variety of problems involving nonstationary states, *provided* that the methods one uses for the calculation of the wavefunctions of all the states of the discrete and of the continuous spectrum are reliable and economic. Such an approach should be capable of utilizing theory and methods from the Quantum Chemistry experience with the *time-independent* Schrödinger equation, including the ab initio calculation of compact wavefunctions for resonance states. For such purposes, an overall many-electron approach is provided by the theoretical framework and methods of the *state- and property-specific theory* of states in the discrete and the continuous spectrum [10, 11].

That research program led the Athens team (Mercouris, Komninos, Nicolaides) to the formulation and proposal of the SSEA, whose first application to the time-dependent ionization of a system with strong electron correlation was published in 1994 [19]. It had to do with the interaction of laser pulses with the ground state of $Li^-$, i.e. with a state of a negative ion where near-degeneracy effects must be included in zero order. The continuous spectrum had two open channels, with thresholds the $Li$ $1s^2 2s$ $^2S$ and $1s^2 2p$ $^2P^o$ states.

Another early demonstration of the versatility of the SSEA and of its contribution to the quantitative understanding of the response of atomic states to strong and short pulses was published in 1996 [20]. It had to do with a problem of time-dependent ionization by strong as well as weak laser pulses, of the $He$ $1s2s$ $^1S$ metastable state, which is the prototypical example of an open-shell excited state that is not the lowest of its symmetry. The applications were concerned with the calculation of a number of observables, such as high harmonic generation, photoelectron angular distributions, etc. The results '*permitted conclusions about the fingerprints left on observable spectra by electron correlations, by Rydberg levels and by doubly excited states*'. (Abstract of [20]).

### 2c. *The non-perturbative theory of the non-linear response of atoms (molecules) to strong fields must account reliably for Rydberg and doubly excited states*

The quoted statement of the last paragraph of *2b* is directly linked to the fact that, when the problem under investigation requires a non-perturbative treatment, the role of the possible participation in the non-linear physics of both Rydberg and doubly excited states (DES) in each channel must be understood quantitatively and with transparency. This is mainly because the non-linearity of the process is intensity-dependent, and therefore, even formally, the accurate contribution of various states to the overall solution is not predictable with the ease with which it is done when lowest-order perturbation theory is applicable. The implication is that the corresponding wavefunctions must be reliable, i.e., they must be as state-specific as possible. For neutrals and negative ions, the DES normally represent resonance states, except when symmetry forbids coupling to the scattering continuum.

The SSEA, whose elements are described in the next subsection, incorporates the state-specific wavefunctions of Rydberg and DES explicitly. Here, in order to illustrate the argument of the first paragraph via a simple example, I recall one of the results from the non-perturbative state-specific calculation of the non-linear response (ionization and polarization) of the ground state of Helium, $1s^2$ $^1S$, as a function of frequency and intensity of the field, published in 2000 [21]. That work implemented the '*many-electron, many-photon*' (MEMP) theory, which was introduced in the late 1980s [22,11]. Its framework is that of unstable states in the continuum, and uses a two-part trial N-electron wavefunction with a combination of separately optimized function spaces with real and complex coordinates. Unlike the '*complex coordinate rotation*' (CCR) method, the coordinates of the Hamiltonian in the MEMP theory remain real [22, 11]. As regards the formal structure of the nonstationary wavefunctions, one or two basic features are similar to those of the SSEA.

Figure 5 of [21] depicts the two-photon ionization rate of He as a function of frequency and of intensity. As the photon frequency reaches the value of ω = 0.74 a.u and the intensity keeps increasing, a peak emerges from the smooth background of the continuum, while the structure due to the intermediate discrete Rydberg state $1s2p$ $^1P^o$ disappears. As regards the appearance of the new peak, we wrote, '*This structure is absent for field intensities of the order $3.5x10^{12}$ W/cm$^2$ or smaller, where the ionization rate increases smoothly and quadratically with intensity*'. It is '*…a manifestation of a higher-order effect, whereby the autoionizing state $2s2p$ $^1P^o$, whose field-free position is 60.2 eV, is coupled resonantly with the He $^1S$ ground state via a three-photon process*' (page 9 of [21]).

In other words, even though in lowest order the two-photon ionization process from the $^1S$ state reaches only $^1S$ and $^1D$ states of even parity, as field-intensity increases the higher-order aspect of dynamics manifests itself via, on the one hand the disappearance of the structure corresponding to the $^1P^o$ Rydberg intermediate discrete state and, on the other hand, the appearance of the lowest $^1P^o$ DES inside the continuous spectrum.

It is important to add here that once the continuous spectrum is involved, additional field-induced couplings for DES of different symmetry and parity may take place. Thus, it has been demonstrated quantitatively how a DES is coupled to another DES, which may be either a discrete state embedded into the continuum or another resonance state. The concomitant phenomenon of field-induced (DC or AC fields) *width variation* was computed non-perturbatively in [23, 24].

In cases such as the ones to which I referred above, experience shows that, unless the wavefunctions of the significant to each problem states are introduced into the calculation explicitly and with numerical accuracy, the predictions of any non-perturbative formalism, (let alone a perturbative one to lowest order), will deviate from the real physics, especially as the intensity of the pulse increases.

As we shall see below, the SSEA to the non-perturbative solution of the METDSE aims at the systematic satisfaction of such requirements.

*2d. The key features of the SSEA*

The physical situations that the present theory covers, (which are numerous), are assumed to be described by a total Hamiltonian (for an atom or a molecule) of the form $H(t) = \mathbf{H}_{atom} + V(t)$. This standard form implies that the formalism can be developed in terms of eigenstates of $\mathbf{H}_{atom}$, and that its computational application to the problem of interest can in principle achieve convergence by invoking the fundamental theorem of quantum mechanics regarding the expansion of a square-integrable function (wavepacket) in terms of the complete set of states of the Hermitian $\mathbf{H}_{atom}$. Obviously, if one considers extreme cases where the strength of the interaction is huge with respect to the forces inside the initial atomic state, the formal and quantitative description of the physical phenomenon using as basis the states of $\mathbf{H}_{atom}$ cannot be justified a priori.

Following the analysis of each problem, the SSEA solution, $|\Psi(t)>_{SSEA}$, is constructed and computed in the form, (I omit the index for each possible channel),

$$|\Psi(t)>_{SSEA} = \sum_m a_m(t)|m> + \int_0^\infty b_\varepsilon(t)|\varepsilon> d\varepsilon \qquad (1)$$

The expansion (1) holds for atoms as well as for molecules. $|m>$ represents the relevant discrete states and $|\varepsilon>$ are energy- normalized (Dirac normalization) scattering states. The mixing complex coefficients are time-dependent. The presence of resonance states in the continuous spectrum of $|\varepsilon>$ is taken into account by adding the corresponding square-integrable component [1, 11].

By its very construction and computational methodology, the SSEA is applicable to many classes of problems where the pulses have wavelengths ranging from, say, the IR to the X-ray regimes, while allowing the direct monitoring of the relative significance of each physical state in eq. 1.

The type and quality of the wavefunctions, as well as the number of the scattering states of the energy-discretized continuum, are determined by each problem of interest, based on results of convergence and depending on the interplay between the characteristics of the radiation pulse and those of the spectrum of the atom (molecule), or of their positive or negative ions.

Substitution of the form (1) into the METDSE transforms it into a system of coupled integro-differential equations, which is solved numerically. For strong fields, the number of these equations ranges from tens to hundreds of thousands, because of the large number of energy-normalized scattering wavefunctions which is normally necessary. As we have stated in previous publications, (e.g., [2,9]), regardless of the number of electrons of the system, the number of the state-specific bound wavefunctions representing the relevant to each problem discrete and resonance states is always small. It is the number of the scattering states of the continuous spectrum, (a different set for each channel), that determines the size of the overall calculation.

Interchannel coupling is taken into account via the solution of the coupled equations using matrix elements of the full Hamiltonian. Once convergence has been obtained, the complex expansion coefficients represent probability amplitudes, provided the length form of the electric dipole approximation is used [2].

As it has been demonstrated in various applications to electronic dynamics, we do not use 'absorbing potentials' to account for wavefunction reflection which may appear as propagation time increases. Instead, since the continuous spectrum is represented by state-specific, energy-normalized scattering orbitals, (computed numerically for the potential of the term-dependent core), the solution of this problem is found by testing, with respect to convergence and to the constraint of conservation of the norm of $|\Psi(t)>_{SSEA}$, the upper limit of the energy spectrum and the density of the energy solutions. For all cases studied thus far, an energy step of 0.001 a.u. has proven sufficient. The time step varies from case to case. The Taylor-series method of time-propagation is explained in our 1994 paper [19].

Provided the coupled equations are solved reliably, (e.g., satisfaction of requirements for numerical accuracy of matrix elements, for propagation as a function of time, etc), the level of accuracy of eq. (1) depends,

i) On the degree of completeness and relevance of the states $|m>$ and $|\varepsilon>$ to the problem under consideration, and,

ii) On the degree to which the stationary trial wavefunctions represent those major components of the exact ones which overwhelmingly determine the diagonal or off-diagonal matrix elements for whose computation they are used.

The above two statements reflect the spirit and the practice of the '*state- and property-specific Quantum Chemistry*' [10].

In addition to the issue of the degree of accuracy of the wavefunctions entering in expansion (1), there is the issue of the calculation of the time-dependent coefficients. For weak fields, it is reasonable to expect that a framework of lowest-order time-dependent perturbation theory (LOTDPT) suffices. For example, by invoking the LOTDPT, a few years ago we derived a time-dependent analytic formula for the time-resolved formation of an isolated resonance upon excitation by a weak and short pulse [8]. This formula employs parameters that characterize the radiation pulse as well as the resonance state. When applicable, its use allows predictions in a rather simple way.

However, for strong fields, the calculation must be non-perturbative, via the solution of the corresponding coupled equations, which is what the SSEA does. In the case of a time-dependent Hamiltonian, $H(t) = \mathbf{H}_{atom} + V(t)$, where an atomic (molecular) state is perturbed by one or more time-dependent radiation pulses with interaction symbolized by $V(t)$, the necessary input for such a SSEA calculation are the state-specific *bound-bound*, *bound-free* and *free-free* matrix elements of $H(t)$, in addition to energies. It must be stressed that, in the non-perturbative regime, the coefficients vary rapidly with the peak intensity of the field, and so the convergence of the overall calculation must be tested carefully.

For many-electron atoms, the *state-specific* wavefunctions for $|m>$ and $|\varepsilon>$ are computed in a manageable way. The scattering functions are computed numerically for each value of the discretized energy above threshold, and for each value of $\ell$, (angular momentum), in a term-dependent Hartree-Fock core potential [2, 10]. Scattering wavefunctions calculated in this way have zero intrachannel coupling. The theory and calculation of the state-specific N-electron wavefunctions are described in [2, 10, 11]. These include the practical treatment of *perturbed* discrete and continuous stationary spectra, in terms of a multi-electron, multi-channel reaction- (K-) matrix theory, whose formulation utilizes function spaces that are suitable for the method of configuration-interaction, the standard tool of Quantum Chemistry, [2, 25, 26].

For diatomic molecules, and especially for polyatomic ones, the non-perturbative solution of the METDSE according to the expansion (1) is a challenging open field, since the reliable and systematic calculation of state-specific excited electronic states, of resonances and of pure scattering states remains a desideratum. For example, for polyatomic molecules there is no rigorous method of numerical integration for the scattering orbitals in the multi-center potential. Of course, for problems that are relatively simple, or are simplified judiciously, expansions which approximate that of eq. (1), (i.e., they approximate the exact time-dependent contribution from each excited state), can always be adopted. For example, such cases may be satisfied by a '*time-dependent multiconfiguration Hartree – Fock*' (TDMCHF)-type or '*time-dependent configuration-interaction*' (TDCI)-type calculation, albeit for special problems only – see below. Indeed, for strong fields and non-perturbative situations, the degree of reliability of approximations to the SSEA will be limited, unless the contribution of the continuous spectrum is taken into account consistently, and in a way which is mathematically justifiable. Here, one must remember that when a molecule is subjected to strong fields, channels of dissociation may compete with those of ionization.

For problems of *multiphoton dissociation or association*, there is no many-electron problem after the construction of the potential energy surface. Hence, the vibrational states of the discrete and the continuous spectrum can be represented after diagonalization, to a good approximation, in terms of suitable analytic basis sets. This approach to the implementation of an approximate SSEA has indeed proven convenient in diatomics [2]. Extension to the more complex problem of time-dependent dissociation of multidimensional potential energy surfaces is possible, once a quantitative understanding of the basis-set representation of all the relevant stationary vibrational states is available. However, when a single set of basis functions is used, the non-perturbative solution of dynamics with more than one open channels becomes complicated.

### *2e. Other computational schemes for the solution of the METDSE for problems of atomic (molecular) states interacting with ultrashort and/or strong pulses*

As new experimental possibilities for the study of phenomena resulting from the interaction of atomic and molecular states with short and/or strong pulses were emerging, the introduction in 1994 of the SSEA [19, 2] was linked to the need at that time to develop theory and *many-electron* methods of computation which are suitable for the nonperturbative solution of the METDSE beyond the cases of one- and two-electron ground states interacting with ultrashort and/or strong pulses. Regardless of the type of physical problem, in pursuing such a goal, the structure of the formalism must allow in a practical way the calculation of effects due to the state-specific features of the pertinent electronic structures and of electron correlations. This requirement is analogous, but much more complex and demanding, to that of many-electron problems of Quantum Chemistry on the energy axis.

Subsequent publications by a number of other groups also present schemes aiming at providing solutions of the METDSE for many-electron systems. Although the scope of the present paper is different, in this subsection I refer briefly to publications where the description and application to many-electron problems of three such schemes can be found. The first is the *time-dependent R-matrix* (TDRM) theory [27-29], a method which was first outlined by Burke and Burke in 1997 [27]. The fact that it was then applied only to a toy model, (multiphoton ionization of a one-dimensional model), is testimony to the difficulties and high demands that such calculations have when it comes to problems with real systems. In a recent publication [29], the TDRM method was applied to a novel problem which is discussed in section 4, namely the determination of the relative *time delay* in the (2*s*,2*p*) photoionization of Neon.

In the other two schemes, methods of Computational Quantum Chemistry are converted into time-dependent ones, in terms of expansions with time-dependent coefficients. One is the TDMCHF, e.g., [30-32], and the other is the TDCI, e.g., [33,34].

Even when the initial-state electronic structure is favorable for their formal implementation, methods such as those in [30-34] express the content of expansion (1) only in an approximate way. They are constrained by the necessary truncation of the basis function space and by the corresponding uncertainties and inaccuracies in the description of the excited states in the discrete and in the continuous spectrum, including Rydberg series (unperturbed or perturbed) and multiply excited resonances. In other words, when it comes to excited states, CI methods using a common basis set produce excited roots which need not constitute a good representation of excited states of the discrete or of the continuous spectrum, especially if the function space of virtual excitations is truncated significantly.

For example, consider the 'TDCI singles' method described in [34]. It is formulated as an expansion over a set of virtual excitations from a zero-order Hartrre-Fock determinant, which includes only one-electron excitations. Yet, even for simple closed-shell initial states, (let alone open-shell, heavily mixed wavefunctions), ignoring the presence of DES, either in the physically significant portion of the spectrum or in the contribution of the character of the excited-state wavefunctions, introduces into the method and into the corresponding results a drastic and uncontrollable uncertainty. Even in the simple case of one-photon transition amplitudes between discrete states, it is known since the 1970s that, although the matter-field interaction is represented by a one-electron operator, orbital relaxation (upon excitation) and electron correlations imply that both singly and doubly (at least) excited configurations must be included in the CI wavefunctions – see the discussion on the '*first-order theory of oscillator strengths*' (FOTOS) in [10]. In fact, this is mandatory when *valence-Rydberg-scattering* state mixing is important in a portion of the spectrum which may contribute to the time-dependent dynamics.

Hence, in spite of the availability of hugely increased computer power, it is a moot question at present whether the structure and computational requirements of these approximate approaches to the solution of the METDSE, can provide accurate solutions to problems of time-resolution of ultrafast electronic processes involving arbitrary electronic structures and spectra, in the context of possible experiments whose parameters are determined by current light sources (e.g., free-electron laser, attosecond pulses). Such experiments may use strong fields, and/or pump-probe schemes with ultrafast pulses, and/or high frequencies which send the system into the continuous spectrum. This means that, for a thorough understanding of the TRMEP, the scattering states cannot be ignored, in addition to contributions from Rydberg and doubly or inner-hole excited discrete or resonance states.

## 3. The first theoretical and experimental results on the details of attosecond time-resolved processes involving electron rearrangements in unstable states

The theme of this section has to do with TRMEP that involves excitation and decay of unstable states, where the role of orbital relaxation and electron correlations is crucial. The first example shows how theory and calculations based on the SSEA had an early encounter with a fundamental problem that later became the first case where the feasibility of the then emerging experimental techniques of '*attosecond metrology*' was demonstrated [35]. This is the problem of the time - resolution of the decay of unstable states in the continuous spectrum [3, 4]. In fact, the resolution achieved by theory via the ab initio solution of the METDSE is more complete than that of the experiment, since it reveals the region of non-exponential decay as well.

The following remarks elaborate on the previous paragraph.

When thinking of, say, an autoionizing state, even though the implied decay is time-dependent, the intrinsic property of energy *width*, $\Gamma$, which is the inverse of the *lifetime* as defined by exponential decay, (ED), is deduced from a measurement on the energy axis, provided the parametrization of the profile is accurate. On the other hand, a fuller picture, with more possible insight, is obtained when the METDSE is solved from first principles, provided a well-defined wavefunction for the square-integrable initial state, $\Psi_0$, has already been computed, and the theory takes into account the rearrangement of the system's electrons as the state dissipates into the continuum of scattering states.

In order to obtain time-resolved information on the decay of real, multielectron unstable states, at very short and at very long times, i.e., outside the realm of ED, in 1996 we introduced the methodology of the SSEA to the calculation of the time-resolved decay of unstable states that are very close to threshold, of neutral atoms and of negative ions, [3, 4]. The cases studied involved processes of barrier penetration for a shape resonance, (e.g., the three-electron negative ion shape resonance $He^- \ 1s2p^2 \ ^4P$), as well as of the two-electron rearrangement which normally dominates non-relativistic or relativistic autoionization.

Among other things, it was demonstrated on real, many-electron unstable systems, that, as the isolated decaying system evolves to very long times, indeed the exponential decay "law" breaks down, a quantitative finding which was, until then, known only as a result of phenomenology from the application of models.

By identifying the ED regime on the decay curve as a function of time, (attosecond-femtosecond scale), it is then possible to extract the lifetime and the corresponding $\Gamma$. For example, for the 20-electron $Ca \ KL3p^63d5p \ ^3F^o$ state, whose bound-free interaction is smooth and nearly constant from zero to about 5.5 eV above its threshold, the lifetime is easily deduced from the exponential decay part of the survival probability to be $3.5 \times 10^{-14}$ sec. [4].

The same information was produced from the first application of experimental attosecond metrology a few years later. Specifically, in 2002, Drescher et al [35], using a novel source of EUV attosecond pulses in conjunction with pump-probe techniques with ultrafast sampling of electron energies, recorded for the first time the time-resolved exponential decay of a $3d$ - hole state in Krypton. As the authors [35] pointed out, the lifetime which they deduced, 7.9 ± 1.0 femtoseconds, corresponds to an Auger width of 84 ± 10 meV, in agreement with a previous measurement in the energy domain, which gave 88 ± 4 meV [36].

The investigation of time-resolved dynamical properties of unstable states using as computational tool the SSEA, was resumed a few years later, as part of the proposal that pointed to the possibility of using future techniques of attosecond spectroscopy for the study of effects of Coulomb strong correlations and electron rearrangements in multiply excited states [5,6]. That work was carried out and published only a few months after the appearance of the two experimental papers reporting the generation of trains [37] or of single [38] attosecond pulses.

Here I recall that, at that time (2001), the focus and achievements of the research efforts were in the realm of the science and technology of the preparation of attosecond pulses. The argument regarding the relevance to possible time-resolved exploration of processes involving electronic 'motions' was vague and intuitive, and was limited to the comment that the revolution of the classically circulating $1s$ electron of Hydrogen takes about 150 *as*. However, this says very little about the physics of real systems, e.g., about the possibility of observable time-dependent, attosecond-scale processes involving states of electrons in atoms or molecules. In this respect, the significant related question must be something like the following: *What kind of time-resolved electron dynamics requires attosecond resolution for it to be observed, and in which systems? And how does theory predict such processes and deal with them quantitatively?*

In 2002, publications [5,6] directed attention to, and provided the first quantitative results for, the connection between resolution on the attosecond scale and effects due to *state-specific SCF orbitals and strong electron correlations* that normally characterize doubly excited and inner-hole excited states. In our conclusions we wrote: '*The results of this study have quantitatively connected the attosecond time-scale to the dynamics of electron correlation in strongly correlated states that is based on the notion of configuration interaction*' (page L278 of [5]). Since then, a number of publications have discussed, and continue doing so, the actual or possible connection of *attosecond spectroscopy* to effects of electron correlations as they manifest in various processes.

Among other things, the work in [5,6] identified strongly correlated Helium DESs as good cases for pump-probe studies of time-resolved electron rearrangements and configuration-interaction at ultrashort timescales. It showed how the coherent excitation and autoionization of three $^1P^o$ DES evolve, while Coulomb '*electron correlation beating*' takes place with amplitudes that are characteristic of the system. Projection onto discrete or scattering stationary states provides the step of observation.

Specifically, the solution of the METDSE was constructed to describe the coherent excitation by two ultra-short pulses and the concurrent decay of DESs of $He$, in transitions from the ground, $1s^2$, or the metastable, $1s2s$, $^1S$ states, to DES of $^1P^o$ symmetry labeled by the $2s2p$, $2p3d$ and $3s3p$ configurations, in the presence of the background of the $1s\varepsilon p$ $^1P^o$ scattering wavefunctions.

Here, it is instructive to demonstrate how, following the analysis of electronic structures in ground and excited states, including resonances, according to the state-specific theory [10,11], the wavefunctions which are used in the SSEA are compact and transparent, yet accurate. As an example, I take the lowest He $^1P^o$ DES discussed above and in [5,6].

As emphasized in the theory of multiparticle resonance states [11], the fundamental requisite for a formally meaningful and computationally tractable theory is to be able to account reliably for the presence of wavefunction localization inside the continuum. In the framework of the state-specific theory, the localized component is computed directly via the numerical solution of the corresponding MCHF equations, and produces the state-specific result,

$$\Psi_0(He\ ^1P^o) = 0.948(2s2p) - 0.308(2p3d) - 0.074(3s3p)$$

The energy obtained with this wavefunction is 60.21 eV above the ground state, when the most accurate experimental value is 60.147 ± 0.001 eV. The corresponding autoionization width is obtained as 0.038 eV, when the experimental value is 0.037 ± 0.001 eV [11].

The framework which allows the self-consistent, state-specific computation of such compact, yet accurate and transparent square-integrable wavefunctions representing the localized component of a multi-electron resonance state, was proposed in 1972, (see the account in [1, 11]), based on a theory of decaying states that emphasized, among other things, the significance of solutions of the many-electron Schrödinger equation with boundary conditions for real as well as for complex energies, and of square-integrability (localization) as the fundamental criterion in the search for the state-specific solution, $\Psi_0$, with a real energy inside the continuous spectrum - see discussion and references in [1,11].

The results and their analysis in [5,6], provided the first insight on time-resolved effects of strong electron correlations that may take place on the attosecond time-scale. For example, it was determined how the correlating superposition of configurations labeling DES in that energy region can lead to ultra-fast intra-atomic rearrangemets of pairs of electrons representing different electronic '*geometries*'. In a simple orbital picture, such a rearrangement concerned the 'motion' of an electron from the *2s* orbital to the *3d* orbital and back, expressed as a superposition of $2s2p$ and $2p3d$ configurations. Assuming that this time-dependent superposition can be probed with attosecond speed, the probability of measuring final states will depend on the dominant configurations, and therefore provide time-resolved signatures of electronic geometry in different states.

Considerations of the type mentioned above [5,6] refer to intra-atomic electronic rearrangements. Obviously, when explored further, they may also prove relevant to experimental investigations aiming at tracing 'motions' of electrons in molecules.

Since 2002, [5,6], aspects of the time-resolved dynamics of states of Helium have attracted the additional attention of various groups, including us, in connection with issues of possible phenomena and observations on ultrashort scales, e.g., [8, 17, 18, 39].

In section 5, I recall the results on the first ab initio demonstration of the time-dependent formation, on the attosecond-femtosecond scale, of electronic resonances in the continuous spectrum represented by doubly excited states of *He* or by inner-hole configurations of the multichannel 13-electron Aluminum.

**4. The application of the SSEA to the calculation of the relative time delay in the (2*s*, 2*p*) photoionization of Neon**

'..*Krausz's more recent work uses attosecond pulses as a probe. In 2010 he and his team discovered that when a 100-eV light pulse impinges on a neon atom, electrons from the atom's 2p orbitals take 21 ± 5 attoseconds longer to be kicked out than do electrons from the 2s orbital [7]. Subtle electron – electron correlations are the delay's most likely cause*'.

C. Day, Physics Today, December 2013 [40]

In order to illustrate the kind of problems which need to be formulated and solved computationally from a many-electron point of view in the advancing science of time-resolved spectroscopy of ultrafast electronic processes, I comment briefly on a novel theme that was introduced and treated in the experimental-theoretical publication [7], to which the Athens team contributed with theoretical concepts and computational results, in collaboration with V. Yakovlev (Garching). The section does not contain any details about the overall experimental and theoretical framework. This information can, of course, be found in [7] and its 'supporting online material' (SOM).

The goal and the corresponding pump (EUV attosecond pulse) – probe (IR laser pulse) '*attosecond streaking*' experiment [7] on the ($2s,2p$) photoionization of Neon were conceived and executed in the 'attophysics' laboratory of the Max Planck institute for Quantum Optics, Garching, Germany. Both the experimental and the theoretical results [7], had to do with an intriguing new question in atomic (molecular) physics: Is it possible that, upon absorption by a many-electron atom of a photon with sufficient energy, electrons from different subshells are emitted with a measurable and computable relative *time delay*?

Although now, a few years after the results and the analysis described in [7], things may look more comprehensible, at the time the experiment was being carried out, (mainly during 2008-9), even the notion of such a time delay in photoionization was absent from the vocabulary of theoretical atomic and molecular physics. Thus, of principal concern was how to identify and compute the quantity that could be related to the measured time delay, and how to understand quantitatively the contributions from basic characteristics of atomic structure, of electron correlation and of transition matrix elements.

The last phrase of the paragraph which is quoted above, indicates the types of issues that had to be explored computationally in order to assure that the quantity of interest is indeed obtained reliably. For example, the scientists who participated in the theoretical project, (Garching, Athens, Vienna), had to consider questions such as:

- How can one formulate and implement an approach for the reliable calculation of a new type of measurable quantity, such as the relative *time delay* in photoionization, and how can he establish the cause(s) of its presence in an experiment such as that described in [7]?
- How can the many-electron nature of the problem be incorporated into the overall approach?
- Can this be done by identifying the bulk of contributions in terms of a zero order model of electronic structure and of time-resolved dynamics, as is often done for more conventional properties and phenomena in atomic and molecular physics?

Questions like the ones above would not be meaningful before the advent of the new science of '*attosecond metrology*' using either trains of attosecond pulses [37 ] or single attosecond pulses [38]. At the same time, they would not be testable and/or predictable computationally within a trustworthy theoretical framework, if it were not possible to solve from first principles, to a physically relevant approximation, the METDSE in terms of the SSEA.

### *4a. Tackling the problem of the relative time delay in (2s,2p) photoionization of Neon [7]*

In the publication [7], the concept of time delay in the photoionization of a many-electron system such as Neon was adopted, albeit not without some concern due to distinct differences, from the theory of particle scattering from a potential of short range, which is associated with the work of Eisenbud, Bohm, Wigner and Smith, e.g., [41]. Although there are details which lead to slightly different definitions and formulations, (e.g., 'dwell' or 'transmission' times, etc), the essence of the concept is that, while in the region of interaction, a moving particle's wavefunction accumulates phase with respect to its hypothetical free wavepacket motion in the absence of the potential. Its mathematical expression involves energy derivatives of phases of complex wavefunctions or of complex transition amplitudes.

For example, even in the simple case of scattering of plane waves, the analysis of Froissart, Goldberger and Watson [42] shows that a time delay in the asymptotic region is given by, $td = \frac{\partial}{\partial \varepsilon_p} \arg f(p, \hat{x} \cdot \hat{p})$, where $f(p, \hat{x} \cdot \hat{p})$ is the scattering amplitude, $\varepsilon_p$ is the energy of the scattered particle and $p$ is the value of the momentum at which the wave packet amplitude is peaked. In the context of the investigation of the time-dependent photoionization of Neon via the SSEA, an analogous expression was used. It involves the energy derivative of the argument of the, state-specific, time-dependent complex coefficients, (eq. 1), of the energy-normalized scattering states (in each channel) that are on resonance with the ionizing pulse. Therefore, the important task is to solve the METDSE and determine the expansion coefficients. The physical assumption for the determination of the relative time delay in the emergence of the two electrons from different subshells is that the effective interaction potentials for each electron are of the order of atomic dimensions and that the zero of time is set at the same instant, the instant when the photon pulse is absorbed.

### *4b. The SSEA calculations*

In the case of the Neon problem, the translation of the previous statements into an ab initio quantum mechanical calculation according to the SSEA was as follows.

In the attosecond streaking experiment [7], the total interaction involves two phase-coherent pulses, the pump attosecond EUV and the probe laser IR, and the measurements are done as a function of the controlled delay, $\Delta t$, between them. The information is recorded in spectrograms that consist of a series of photoelectron energy spectra as a function of $\Delta t$. In other words, in order to deduce the sought after quantity, the experiment adds another parameter, that of the probe, which allows the measurement to be made. The experimental probe was a "*near-single-cycle*" pulse of near-infrared laser, phase-locked with the attosecond pulse, with carrier wavelength of ~ 750 nm ($h\nu$ = 1.65 eV) and duration ~ 3.3 femtoseconds (*fs*). In fact, as part of the overall investigation of this phenomenon, Yakovlev et al [43] showed that "*attosecond streaking enables the measurement of quantum phase*", a result which is directly linked to the analysis presented in [7] and in its SOM.

In principle, the quantum mechanical simulation of the above type of dynamical system is described by the METDSE whose Hamiltonian is expressed as $H_{tot}(t) = H_{Ne} + V_{int}(t)$, where $H_{Ne}$ is the Hamiltonian of the field-free ten-electron Neon atom and $V_{int}(t) = V_{IR}(t) + V_{EUV}(t - \Delta t)$. As is normal, the practical choice for the atom-field coupling operator is one of the forms of the electric dipole approximation.

The (2*s*,2*p*) photoionization of Neon by the attosecond pulse centered at about 106 eV with a width of about 14 eV, implies the dipole transitions $2s \rightarrow \varepsilon p$, $2p \rightarrow (\varepsilon s, \varepsilon d)$ and final states in the continuum of $^1P^o$ symmetry. In the presence of $V_{IR}(t)$, states with many angular momenta in the continuum are mixed. These states are labeled as, $\psi_{2s}^{\ell,L}(\varepsilon) = (1s^2 2s 2p^6)\varepsilon\ell \, ^1L$ and $\psi_{2p}^{\ell',L'}(\varepsilon) = (1s^2 2s^2 2p^5)\varepsilon\ell' \, ^1L'$. The bound wavefunctions of the ground state, $\Phi_{gs}(^1S)$, and of the ionic cores, are computed by including the few single and double orbital electron correlations which contribute with some weight. As explained in [10], in this case, this was done by solving numerically the corresponding MCHF equations. Of significance in the channel $2s \rightarrow \varepsilon p$, is the '*hole-filling*' substitution, $2p^2 \leftrightarrow 2s3d$ [10-12].

If we were to use the SSEA with the $H_{tot}(t)$ above, the corresponding $\Psi_{SSEA}(t)$ would be constructed in the following form:

$$\Psi_{SSEA}(t) = a_{gs}(t)\Phi_{gs}(^1S) +$$

$$+ \sum_{\ell,L}^{\ell=\ell_{max}} \int^{\varepsilon_{max}} b_{2s}^{\ell,L}(\varepsilon,t)\psi_{2s}^{\ell,L}(\varepsilon)d\varepsilon + \sum_{\ell',L'}^{\ell'=\ell'_{max}} \int^{\varepsilon_{max}} b_{2p}^{\ell',L'}(\varepsilon,t)\psi_{2p}^{\ell',L'}(\varepsilon)d\varepsilon \qquad (2)$$

According to the SSEA methodology, the upper limits of the above summations, $\ell_{max}$ and $\ell'_{max}$, as well as the upper limits of the integrals $\varepsilon_{max}$, are varied until full convergence of the time-dependent complex coefficients is achieved.

Following the assertion that treats photoionization as a half-collision process, and the known results of scattering theory, e.g., [41, 42], the time delay for each channel is obtained as $\hbar \frac{d}{d\varepsilon}\arg[b_{2s}^{1,1}(\varepsilon,t)]$ and $\hbar \frac{d}{d\varepsilon}\arg[b_{2p}^{2,1}(\varepsilon,t)]$, in the energy region fixed by the EUV excitation energy. For Neon, as well as for other systems, the quantitative understanding of the degree of sensitivity of these quantities (energy derivatives of phases) on the wavefunctions and on the resulting complex coefficients is a crucial element of theoretical work in this new area of research.

The Hamiltonian used in the SSEA work in [7] contained only the interaction with the ionizing attosecond EUV pulse. Consequently, the number of scattering wavefunctions in eq. (2) is reduced drastically and so the overall calculation was tractable. The justification for this approach is as follows.

The EUV part is the one defining the ionization process. It has all the features of a many-electron problem. For example, questions such as the following arise: How does each electron "feel" its state-specific environment both at the instant of the photon absorption and as it emerges free from the interaction region of the atom? How and to what extent does the correlation with, and rearrangement of, the remaining electrons affect the quantity associated with time delay?

Conversely, the second interaction term, $V_{IR}(t)$, is essentially linked to a one-electron problem in the continuum, albeit in a many-electron potential of the atomic ion which is also state-specific. Here, the serious computational difficulty comes from the field-induced coupling inside the infinitely degenerate continuous spectrum, where, in principle, both on-resonance and off-resonance matrix elements between scattering states must be considered. It should be added that it is reasonably assumed that the bound orbitals of the initial and final states are not affected by the presence of the IR field at the intensities of the experiment ($\approx 10^{11}$ W/cm$^2$).

The solutions of the METDSE for each ionization channel using only $V_{EUV}(t)$, can indeed be used to determine the relative time delay in photoionization. However, the question which arises is whether such a result can be compared directly to the one deduced from the streaking measurements. In this regard, I comment as follows:

*i*) Because of its structure, and contrary to the so-called 'grid methods', there is no problem for the SSEA to handle the case of the two-pulse interaction, $V_{IR}(t) + V_{EUV}(t-\Delta t)$, for a many-electron atom, initially in a state which is described in zero order either by a closed- or by an open-shell configuration of any symmetry. In fact, in different contexts, where the problems are computationally less demanding, the METDSE was indeed solved for a system where the initial state is open shell and interacts with two ultrashort pulses, e.g., [5, 6]. However, in the present case of Neon, implementing the SSEA with both the EUV and the IR pulses would require the calculation and processing of a huge number of free-free coupling matrix elements and a considerable increase (in the hundreds of thousands) in the number of the coupled equations necessary for testing convergence. Neither a few years ago nor today is such a computation practical for us, given the rather limited computational facilities of our institute in Athens.

*ii*) The theoretical framework in [7] indeed used a model for the simulation of the attosecond streaking experiment including the effect of the IR pulse. This was done admirably by Yakovlev and his collaborators in Garching. The model was based on the Coulomb-Volkov approximation (CVA).

As regards the final value of the relative time delay, its contribution essentially did not change the result for the time delay (6.4 *as*) obtained from the calculation that was based on the SSEA alone [7]. This is in harmony with the fact that, for the IR laser intensities used in [7], around $10^{11}$ W/cm$^2$, the experimental results did not vary with changes of the intensity. In other words, it seems that, in this case, the probe did not have a crucial effect on the relative time delay.

The SSEA calculations of 2010 [7] using $V_{EUV}(t)$ took into account,

  i) the pulse characteristics,

  ii) the state-specific nature of orbital electronic structures,

  iii) the significant part of electron correlation in initial and final states, following the state-specific theory reviewed and explained in [10].

  iv) the interchannel coupling.

These calculations were followed by the first investigation of possible effects on the (2*s*,2*p*) time delay of inner- hole doubly excited resonance states of Neon, whose electronic structures, energies and widths were predicted in [44] – see below.

Contrary to our initial anticipation, the results for this system did not show a strong dependence on details of electron correlation beyond the MCHF level, or on interchannel coupling. For example, whereas the final result due to the pump step was 6.4 *as*, the result without interchannel coupling, i.e., using just the dipole matrix elements connecting the initial state-specific orbitals, calculated at the MCHF level, to their corresponding final state channel, was 4 *as*. (See page 1661 of [7]). Similarly, the CVA scheme did not yield significant time-delay for the probe step. I quote from page 6 of the SOM of [7]:

*'To conclude this section, let us illustrate the above result. The matrix elements obtained by means of the state-specific expansion approach (SSEA) predict a(2p) – a(2s) = 6.37 attoseconds when the spectral averaging is performed with a 182-attosecond XUV pulse with a central photon energy of 106 eV. The spectrograms simulated with the aid of the CVA exhibit a temporal shift of 6.38 attoseconds'*. (SOM of [7]).

The theoretical results corroborated the essence of the experimental findings. Specifically, it was demonstrated that, indeed, there is *time-delay* in atomic photoionization, in the following sense: Upon absorption of the photon, assumed to occur instantaneously, the emergence of the wavepackets (of different kinetic energies) representing the electrons initially occupying the Neon 2*p* and 2*s* Hartree-Fock orbitals takes place with a relative delay. In agreement with experiment, the '2*p* electron' emerges a few attoseconds after the '2*s* electron' does.

On the other hand, whereas the analysis of the measurements led to the conclusion that the photoionization relative time delay for Neon (2*s*,2*p*) is 21 ± 5 *as*, the analysis of the theoretical results produced a number of 6.4 *as*.

In search of an explanation of this discrepancy, we turned to the possibility that the presence of resonance states in the energy region of interest, (excitation energy of about 106 eV), might be responsible. At that time, no such resonance states were known to exist in this energy region, either theoretically or experimentally, and hence none had been included in the original calculations. The search for such a possible explanation, led us to the identification and calculation via the state-specific polyelectronic theory of unstable states explained in [11] of a series of a new type of DES in the Neon spectrum. These states are labeled by $2s^2$ –hole configurations and are embedded in double electron continua. One of them has its energy in exactly the right region [44]. However, because of its narrow width, upon spectral averaging it does not contribute to the time delay in this problem. In the conclusion of [44], one reads:

"*Finally, we comment on the possible relevance of the present results to a recently published experimental-theoretical study of delay in photoemission [7]. That paper investigated the delay in emission between the 2s and the 2p electrons of neon upon the absorption of a photon pulse with the experimental energy of 106 eV and a width at half maximum of 14 eV. The fact that the herein predicted first $^1P^o$ resonance at 105.9 eV is so close to the photon energy of Ref. [7] has motivated us and a colleague (V. Yakovlev) to explore the possibility that the presence of this resonance influences the overall time delay as measured in Ref. [7]. However, calculations using the approach described in Ref. [7] showed that this is not the case due to the fact that the width of this resonance is very narrow*".

The enigma stemming from the reported difference between the experimental and the theoretical findings in [7] remains to date, in spite of subsequent theoretical investigations by a number of researchers. It is challenging to both theory and experiment. In this context, it is worth noting that the same theoretical number, i.e., 6.4 *as*, was reported recently by Feist et al [45] who carried out very large R-matrix type calculations, including a series of resonance states near and far from the critical region of 106 eV. Their results confirm the original theoretical predictions which were made in [7, 44].

I close this section by pointing out that, following the publication of the papers [7] and [44], and their demonstration, among other things, of the quantitative relevance of energy derivatives of phases and of many-electron calculations, including the consideration of interchannel coupling and of resonances, a number of theoretical papers have reported results on the theme of time delay in photoionization. I distinguish the discussions and computations of the Belfast [29] and of the Vienna teams, [15, 45, 46]. The latter also contributed to [7]. Other papers can be found in the reference lists of [15, 45, 46].

## 5. Time-dependent formation of the excitation profile of an autoionizing state on ultra-short time scales

The 2002 analysis and calculations based on the SSEA [5,6] aimed at demonstrating that one area of possible application of a future ultrafast time-resolved spectroscopy on the attosecond scale is that of the dynamics of the coherent excitation and decay (autoionization) of strongly correlated unstable states in the continuous spectrum. The prototypical system which was studied from first principles was Helium, excited to $^1P^o$ doubly excited resonance states during the photoionization of either the first excited singlet state, $1s2s\ ^1S$, or the ground state, $1s^2\ ^1S$. In order to drive the excitation efficiently, two femtosecond pulses were used in the time-dependent Hamiltonian.

In 2005, Wickenhauser et al [47], using a model system which assumes data from (super)Coster-Kronig transitions with lifetimes of ~ 400 attoseconds, explored an aspect of the phenomenology of autoionizing states which is novel and provides insight into the physics of electron correlations in the continuous spectrum on the time axis. They "*investigated the feasibility of observing the buildup of a Fano resonance in the time domain by attosecond streaking techniques*", (abstract of [47]).

Their publication led us to return to the problem of the ab initio solution of the METDSE for the Helium system in order to produce, for the first time, quantitative results for the time-dependent formation of the asymmetric profile of a resonance state, i.e., of the $He\ 2s2p\ ^1P^O$ DES, upon excitation by a well-specified ultra-short pulse [8]. This entailed the implementation of the SSEA to determine how the effects of the correlations of the pair of the initially bound electrons evolve and eventually form the stationary superposition of bound with scattering components on the energy axis that gives rise to the asymmetric Fano profile. In addition, using first-order time-dependent perturbation theory, we derived and applied an easily applicable analytic formula for this phenomenon, valid for weak pulses. It was shown that, for weak fields, its results indeed coincide with those obtained from the non-perturbative solution of the METDSE (Figure 2 of [8a]).

**Figure 1** shows the time-resolved differential ionization probability, $P(\varepsilon,t)$, for the gradual formation of the asymmetric profile of the of the $He\ 2s2p\ ^1P^o$ resonance state upon excitation from the ground state by a $\sin^2$ pulse of duration 450 a.u (11 fs) and of energy of about 60.1 eV. (The figure is taken from our work in [8]). $P(\varepsilon,t)$ is defined by the projection of $\Psi_{SSEA}(t)$ on the scattering stationary state at energy $\varepsilon$, $P(\varepsilon,t) \equiv |< scattering\ state(\varepsilon)|\Psi_{SSEA}(t) >|^2$, where $\varepsilon$ is the energy above the ionization threshold. As the two electrons in the bound orbitals $2s$ and $2p$ are allowed to correlate, and localized as well as asymptotic components are added, the quantity $|< \Psi_0(2s2p)|\mathbf{H}|1s\varepsilon p >|^2 d\varepsilon$ dominates the transition of the two-electron wavepacket to the continuum of the scattering states in the region around $E_0$, while interference effects eventually stabilize after about 180 fs, when the completion of the formation of the stationary superposition of the resonance takes place.

It is important to add that the position, the width and the Fano *q* parameter which were determined for $He\ 2s2p\ ^1P^o$ from the time-dependent SSEA calculation, are in perfect agreement with the same quantities that have been measured on the energy axis [8].

As a proof-of-principle demonstration of the same theory and computational methodology to a many-electron, multi-channel system, in 2007 we also published results, similar to those of the previous applications to Helium, on the time-resolved formation of autoionization states in the multi-channel continuum of the 13-electron Aluminum [9].

I close by pointing out that the analytic formula derived (corrected for a misprint) and used in [8], can be used for appropriate predictions in many-electron atoms, provided its parameters are known from computation or experiment. For example, its application shows that Auger states of atoms in the middle of the Periodic Table, whose width is sufficiently large, about 2 eV or more, are suitable candidates for observing the formation of their stationary profile with attosecond duration.

## 6. Epilogue

As explained in the previous sections and their references, during the past two or three decades new vistas of research have opened in experimental and theoretical *time-resolved many-electron physics* (TRMEP) of ultrafast electronic processes, indicating the presence of a new era of Quantum Chemistry and the many-electron problem [1].

These developments have to do with the science of the interaction of strong and/or ultrashort electromagnetic pulses with atomic (molecular) states. The related problems most often involve both the discrete and the continuous electronic spectrum. Their understanding and reliable solution, for purposes of interpretation as well as of prediction of results, have as a fundamental prerequisite the possibility of computing non-perturbatively the appropriate for each system (atom plus light pulse(s)) solution, $\Psi(t)$, of the *many-electron time dependent Schrödinger equation* (METDSE), for which the total Hamiltonian is $H(t) = \mathbf{H}_{Atom} + V(t)$. The interaction operator, $V(t)$, includes the characteristics, (field strength, duration, shape), of the pulse (or of the pulses) used in the problem.

A general framework and related methodology for the solution of the METDSE is described by the *state-specific expansion approach* (SSEA), [2,10], which was outlined in section 2.

According to the SSEA, it is feasible and rewarding to search for the solution of the METDSE for various classes of problems, by capitalizing on the theory and the computational experience concerning the quantitative understanding of the variety of electronic structures of stationary states in both the discrete and the continuous spectrum. For example, when it was introduced in 1994 [19], the many-electron problem which was solved was that of the time-dependent multiphoton ionization of a strongly correlated, multiconfigurational ground state, with a multichannel continuum. Specifically, we chose a negative ion, $Li^-$, and the initial state was labeled by an MCHF wavefunction, $0.93\ (1s^2 2s^2) + 0.36(1s^2 2p^2)\ ^1S$. The final channels were defined by the $1s^2 2s\ ^2S$ and $1s^2 2p\ ^2P^o$ states of $Li$. The solution of the METDSE for states of negative ions with near-degeneracies constitutes a good test for the quality of any formalism and methodology.

The central argument of the SSEA is to invoke the fundamental equation of Quantum Mechanics concerning the expansion of $\Psi(t)$ over the stationary states of the Hamiltonian. Hence, by taking into account the characteristics of each problem, the foundational starting point is the construction of expansions such as those of eqs. (1) and (2). This means that the N-electron stationary wavefunctions representing discrete, scattering and resonance states must be as *state-and property-specific* as possible [2,10]. For example, this is done straight-forwardly, conceptually and practically, for the hydrogen atom where the state-specific eigenfunctions of the discrete and the continuous spectrum are known exactly. (Of course, no resonance states are present in the hydrogen spectrum). However, when it comes to many-electron arbitrary electronic structures and complex spectra, one is obviously faced with multifaceted problems and serious challenges. Some of them have already been treated. Obviously, many remain, especially as regards the (approximately) state-specific calculation and use of excited wavefunctions in polyatomic molecules.

The critical element in the solution of the coupled equations of the SSEA is the correct and expedient calculation of the three sets of matrix elements of $H(t)$,

'*bound-bound*', '*bound-free*', and '*free-free*', for a given form of the coupling operator. The atomic or molecular electronic Hamiltonian can be either non-relativistic or relativistic. In practice, appropriate approximations are applicable, based on theory and analysis of electronic structures and matrix elements, as described in [2,10]. For example, for unperturbed Rydberg states with a core where near-degeneracies are absent, when the coupling operator is the length form of the dipole approximation, one must concentrate on the numerical accuracy of the outer orbital in the fixed, term-dependent Hartree-Fock core. Conversely, for those multiply excited discrete or resonance states that is deemed necessary to be included in the expansion, their wavefunctions must be computed by taking into account the dominant components representing electron correlations.

The examples which were discussed briefly in sections 3-5 represent prototypical cases for which, by implementing the SSEA, theory has produced quantitative explanations and predictions for time-resolved effects of state superpositions, and of electron correlations and rearrangements on femtosecond and attosecond scales. I point to two of these examples:

The first is the quantitative prediction from first principles of the time-dependent formation of the profile of resonance states in Helium (doubly excited) and in Aluminum, (inner-hole excited), as bound-scattering interference is building up and is completed (Figure 1 and refs. [8,9]). This work moved along the path paved in [5,6], where concrete proposals and results concerning the connection of attosecond spectroscopy to time-resolved dynamics dictated by state-specific orbital configurations and correlations were presented.

The second is the new type of TRMEP on the attosecond scale discussed in section 4 based on [7]. The work published in [7] on the relative *time delay* in the photoionization of Neon provided novel insight, experimentally as well as theoretically, from a time-dependent point of view, on a phenomenon that has been around for a century, but which, until 2010, had always been treated on the energy axis.

**Caption for Figure 1**

Time-resolved differential ionization probability, $P(\varepsilon,t)$, for the gradual formation of the asymmetric profile of the $He\ 2s2p\ ^1P^o$ resonance state upon excitation from the ground state by a $\sin^2$ pulse of duration 450 a.u (11 fs). $\varepsilon$ is the energy above the ionization threshold. After about 180 femtoseconds, the characteristics of the curve, (the position, the Fano *q* parameter and the width), are the same as the ones that have been obtained from various treatments on the energy axis. (From [8]. Copyright (2007) by the American Physical
Society).

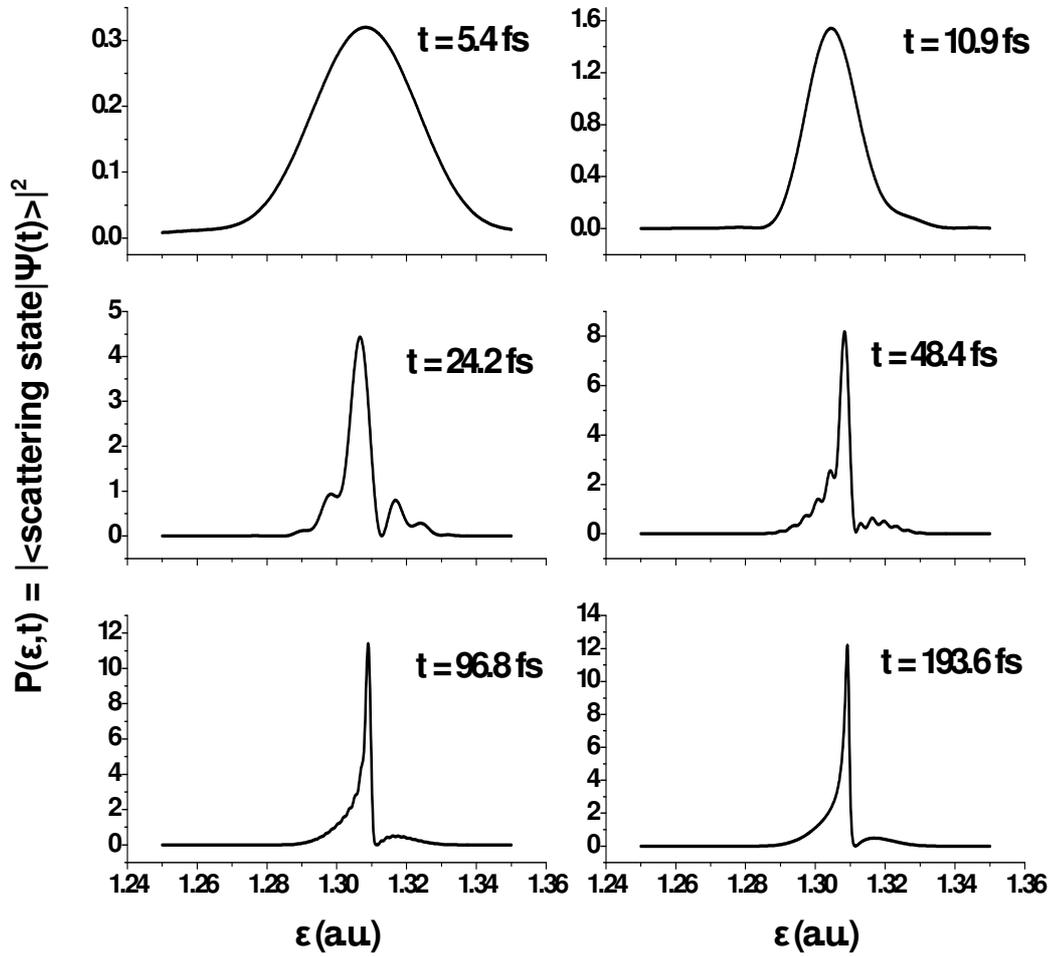